# A Vector-Based Representation of the Chemical Bond for the Substituted Torsion of Biphenyl

**Jiahui Li, Weijie Huang, Tianlv Xu, Steven R. Kirk[*] and Samantha Jenkins[*]**

*Key Laboratory of Chemical Biology and Traditional Chinese Medicine Research and Key Laboratory of Resource Fine-Processing and Advanced Materials of Hunan Province of MOE, College of Chemistry and Chemical Engineering, Hunan Normal University, Changsha, Hunan 410081, China*

*email: samanthajsuman@gmail.com
*email: steven.kirk@cantab.net

We use a new interpretation of the chemical bond within QTAIM, the bond-path framework set $\mathbb{B} = \{p, q, r\}$ with associated linkages with lengths $\mathbb{H}^*$, $\mathbb{H}$ and the familiar bond-path length is used to describe a torsion θ, $0.0° \leq θ < 22.0°$ of *para*-substituted biphenyl, $C_{12}H_9\text{-}x$, $x = N(CH_3)_2, NH_2, CH_3, CHO, CN, NO_2$. We include consideration of the H---H bonding interactions and find that the lengths $\mathbb{H} > \mathbb{H}^*$ that we explain in terms of the most and least preferred directions of charge density accumulation. We also consider the fractional eigenvector-following path lengths $\mathbb{H}_f$ and $\mathbb{H}_{fθmin}$.

A *para*-substituted biphenyl molecule subjected to an applied torsion θ provides a useful method to probe a new interpretation of the chemical bond that we refer to as the bond-path framework set $\mathbb{B}$[1]. This is due to the presence of both the strong torsion bond that links the phenyl rings and the associated weak H---H bonding interactions. Recently, some of the current authors used the quantum theory of atoms in molecules (QTAIM)[2] and the stress tensor topological approaches to add to the long standing discussion about both the biphenyl[3] and substituted biphenyl molecule[4,5]. We explained the effects of the torsion θ of the C-C bond linking the two phenyl rings of the biphenyl molecule and demonstrated the favorable conditions for the formation of the H---H bonding interactions for a planar biphenyl geometry. This QTAIM analysis was found to be consistent with an earlier alternative QTAIM[6] approach that indicated that the H---H bonding interactions provided a locally stabilizing effect that was overwhelmed by the destabilizing effect of the C-C bond. The H---H bonding interactions exist in a wide range of chemical environments[7] including biphenyl[8,9].

The H---H *BCP*s are very tenuous interactions and therefore can be used as a sensitive probe to detect subtle differences in the chemical environment in the context of the new QTAIM vector-based interpretation of the chemical bond, referred to as the *bond-path frame-work set* $\mathbb{B}$. An important goal for this work therefore, is to determine the effect the applied torsion θ has on the C-C torsion bond and H---H bonding due to the six *para*-substituted groups of $C_{12}H_9$-x where x = $N(CH_3)_2$, $NH_2$, $CH_3$, CHO, CN, $NO_2$ as quantified by $\mathbb{B}$.

We use a new QTAIM analysis that utilizes higher derivatives of $\rho(\mathbf{r_b})$ in effect, acting as a 'magnifying lens' on the $\rho(\mathbf{r_b})$ derived properties of the wave-function. We will use QTAIM[2] to identify critical points in the total electronic charge density distribution $\rho(\mathbf{r})$ by analyzing the gradient vector field $\nabla\rho(\mathbf{r})$. These critical points can further be divided into four types of topologically stable critical points according to the set of ordered eigenvalues $\lambda_1 < \lambda_2 < \lambda_3$, with corresponding eigenvectors $\underline{\mathbf{e}}_1$, $\underline{\mathbf{e}}_2$, $\underline{\mathbf{e}}_3$ of the Hessian matrix. The Hessian of the total electronic charge density $\rho(\mathbf{r})$ is defined as the matrix of partial second derivatives with respect to the spatial coordinates. The eigenvector $\underline{\mathbf{e}}_3$ indicates the direction of the bond-path at the *BCP*. The most and least preferred directions of electron accumulation are $\underline{\mathbf{e}}_2$ and $\underline{\mathbf{e}}_1$, respectively[10–12]. The ellipticity, ε provides the relative accumulation of $\rho(\mathbf{r_b})$ in the two directions perpendicular to the bond-path at a *BCP*, defined as $\varepsilon = |\lambda_1|/|\lambda_2| - 1$ where $\lambda_1$ and $\lambda_2$ are negative eigenvalues of the corresponding eigenvectors $\underline{\mathbf{e}}_1$ and $\underline{\mathbf{e}}_2$ respectively.

The four types of critical points are labeled using the notation (R, ω) where R is the rank of the Hessian matrix, the number of distinct non-zero eigenvalues and ω is the signature (the algebraic sum of the signs of the eigenvalues); the (3, -3) [nuclear critical point (*NCP*), a local maximum generally corresponding to a nuclear location], (3, -1) and (3, 1) [saddle points, called bond critical points (*BCP*) and ring critical points (*RCP*), respectively] and (3, 3) [the cage critical points (*CCP*)]. In the limit that the forces on the nuclei become vanishingly small, an atomic interaction line[13] becomes a bond-path, although not necessarily a chemical bond[14]. The complete set of critical points together with the bond-paths of a molecule or cluster is referred to as the molecular graph.

The bond-path length (BPL) is defined as the length of the path of the **e**$_3$ eigenvector of $\lambda_3$ eigenvalue, defined at the *BCP*, of the Hessian of the total charge density $\rho(\mathbf{r})$ that follows the maximum in $\rho(\mathbf{r})$. The bond-path curvature separating two bonded nuclei is defined as the dimensionless ratio:

(BPL - GBL)/GBL  (1)

Where the bond-path length (BPL) is defined as the length of the path traced out by the **e**$_3$ eigenvector of the Hessian of the total charge density $\rho(\mathbf{r})$, passing through the *BCP*, along which $\rho(\mathbf{r})$ is locally maximal with respect to any neighboring paths. Where BPL defined to be the bond-path length associated and GBL is the inter-nuclear separation. The BPL often exceeds the GBL particularly in for weak or strained bonds and unusual bonding environments[15].

We trace out the length of the path swept by the tips of the scaled **e**$_2$ eigenvectors of the $\lambda_2$ eigenvalue, using the ellipticity $\varepsilon$ as the scaling factor. With $n$ scaled eigenvector **e**$_2$ tip path points $q_i = r_i + \varepsilon_i \mathbf{e}_{2,i}$ on the path $q$ where $\varepsilon_i$ = ellipticity at the $i^{th}$ bond-path point $r_i$ on the bond-path $r$, see equation (**2a**). It should be noted that the bond-path is associated with the $\lambda_3$ eigenvalues of the **e**$_3$ eigenvector does not take into account differences in the $\lambda_1$ and $\lambda_2$ eigenvalues of the **e**$_1$ and **e**$_2$ eigenvectors. Analogously, equation (**2b**), is used for the **e**$_1$ tip path points we have $p_i = r_i + \varepsilon_i \mathbf{e}_{1,i}$ on the path $p$ where $\varepsilon_i$ = ellipticity at the $i^{th}$ bond-path point $r_i$ on the bond-path $r$.

We will refer to the new QTAIM interpretation of the chemical bond as the *bond-path framework set* that will be denoted by $\mathbb{B}$, where $\mathbb{B} = \{p, q, r\}$. This effectively means that in the most general case a bond is comprised of three 'linkages'; $p$, $q$ and $r$ associated with the **e**$_1$, **e**$_2$ and **e**$_3$ eigenvectors respectively.

From this we shall define eigenvector-following path length $\mathbb{H}^*$ and $\mathbb{H}$, see **Scheme 1**:

$$\mathbb{H}^* = \sum_{i=1}^{n-1} |p_{i+1} - p_i| \quad (2a)$$
$$\mathbb{H} = \sum_{i=1}^{n-1} |q_{i+1} - q_i| \quad (2b)$$

The *eigenvector-following path* length $\mathbb{H}^*$ or $\mathbb{H}$ refers to the fact that the tips of the scaled **e**$_1$ or **e**$_2$ eigenvectors will sweep out along the extent of the bond-path, defined by the **e**$_3$ eigenvector, between the two bonded nuclei that the bond-path connects. In the limit of the value of the ellipticity $\varepsilon = 0$ *for all* steps $i$ along the bond-path then $\mathbb{H}$ = BPL and for the more general case $\mathbb{H}$ > BPL.

From the form of $p_i = r_i + \varepsilon_i \mathbf{e}_{1,i}$ and $q_i = r_i + \varepsilon_i \mathbf{e}_{2,i}$ we see for shared-shell *BCP*s, that in the limit of the ellipticity $\varepsilon \approx 0$ i.e. corresponding to single bonds, we then have $p_i = q_i = r_i$ and therefore the value of the lengths $\mathbb{H}^*$ and $\mathbb{H}$ attain their lowest limit; the bond-path length ($r$) BPL. Conversely, higher values of the ellipticity ε, for instance, corresponding to double bonds will always result in values of $\mathbb{H}^*$ and $\mathbb{H}$ > BPL.

The bond-path framework set $\mathbb{B} = \{p, q, r\}$ considers the bond-path to comprise the *unique* paths, *p*, *q* and *r*, swept out by the $\underline{e}_1$, $\underline{e}_2$ and $\underline{e}_3$, eigenvectors that form the eigenvector-following path lengths $\mathbb{H}^*$, $\mathbb{H}$ and BPL respectively. The paths *p* and *q* are unique even when the lengths of $\mathbb{H}^*$ and $\mathbb{H}$ are the same or very similar because *p* and *q* traverse different regions of space.

Non-linear bond-paths as determined by equation **(1)** are more likely to occur for the equilibrium geometries of closed-shell *BCP*s rather than for shared-shell *BCP*s. Additionally, confined and symmetrical (e.g. H---H *BCP*) as opposed to 'dangling' (C-H *BCP*) bond-paths *r* with non-zero bond-path curvature will usually result in the length $\mathbb{H}$ > $\mathbb{H}^*$, see **Scheme 2**. The preference for the distortion of curved bond-paths in the $\underline{e}_2$ direction was observed earlier by one of the current authors. It was found for long weak closed-shell *BCP*s such as O---O *BCP*s that the excess length of a non-linear bond-path was due to the bond-path preferentially bending to a much great extent in the preferred direction $\underline{e}_2$ compared with the least preferred direction $\underline{e}_1$[16]. This preferential bending of the H15---H20/H14---H17 *BCP* bond-paths in the preferred direction $\underline{e}_2$ is demonstrated in **Scheme 2**.

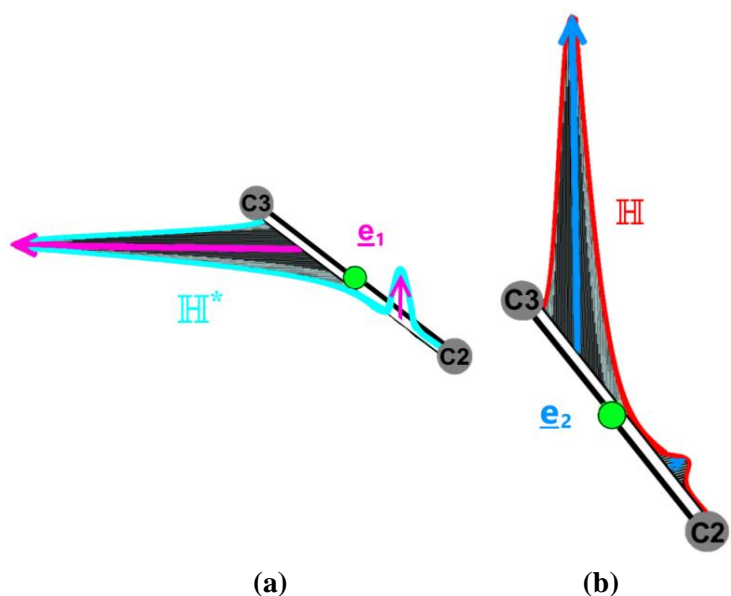

(a) (b)

**Scheme 1.** The pale blue line in sub-figure **(a)** represents the path, referred to as the eigenvector-following path length $\mathbb{H}^*$, swept out by the tips of the scaled $\underline{e}_1$ eigenvectors, shown in magenta, and defined by equation **(2a)**. The red path in sub-figure **(b)** corresponds to the eigenvector-following path length $\mathbb{H}$, constructed from the tips of the scaled $\underline{e}_2$ eigenvectors, shown in mid-blue and is defined by equation **(2b)**. The pale and mid-blue arrows representing the $\underline{e}_1$ and $\underline{e}_2$ eigenvectors are scaled by the ellipticity ε respectively, where the vertical scales are exaggerated for visualization purposes. The green sphere indicates the position of a given *BCP*. Details of how to implement the calculation of the eigenvector-following path lengths $\mathbb{H}^*$ and $\mathbb{H}$ are provided in the **Supplementary Materials S1**.

Analogous to the bond-path curvature, see equation **(1)**, we may define dimensionless, *fractional* versions of the eigenvector-following path length $\mathbb{H}$ where several forms are possible and not limited to the following:

$\mathbb{H}_f = (\mathbb{H} - \text{BPL})/\text{BPL}$             **(3a)**

$$\mathbb{H}_{f\theta min} = (\mathbb{H} - \mathbb{H}_{\theta min})/\mathbb{H}_{\theta min} \qquad (3b)$$

Where $\mathbb{H}_{\theta min}$ is defined as the length swept out by the scaled **e₂** eigenvectors using the minimum value of the torsion θ, with similar expressions for $\mathbb{H}^{*}_{f}$ and $\mathbb{H}^{*}_{f\theta min}$ can be derived using the **e₁** eigenvectors. The form of $\mathbb{H}_f$ defined by equation **(3a)** is the closest to the spirit of the bond-path curvature, equation **(1)**.

The bond-path framework set $\mathbb{B}$ is constructed from the orthogonal triad of the **e₁**, **e₂** and **e₃** eigenvectors and therefore will be suitable for capturing the torsional motion of the *para*-substituted biphenyl molecule $C_{12}H_9$-*x* where *x* = $N(CH_3)_2$, $NH_2$, $CH_3$, CHO, CN, $NO_2$. We can compare the effect on the {**e₁**, **e₂**, **e₃**} bond-path frameworks of the shared-shell torsional C4-C7 *BCP* and closed-shell H---H *BCP*s, for all choices of the substituent *x*, see **Scheme 2**. This will be undertaken in terms of the bond torsion and stretching/compression of the bond-path framework set $\mathbb{B}$; $\mathbb{H}^{*}$, $\mathbb{H}$ and BPL respectively as well as the corresponding fractional measures ($\mathbb{H}^{*}_{f}$,$\mathbb{H}_{f}$), ($\mathbb{H}^{*}_{f\theta min}$,$\mathbb{H}_{f\theta min}$) and bond-path curvature.

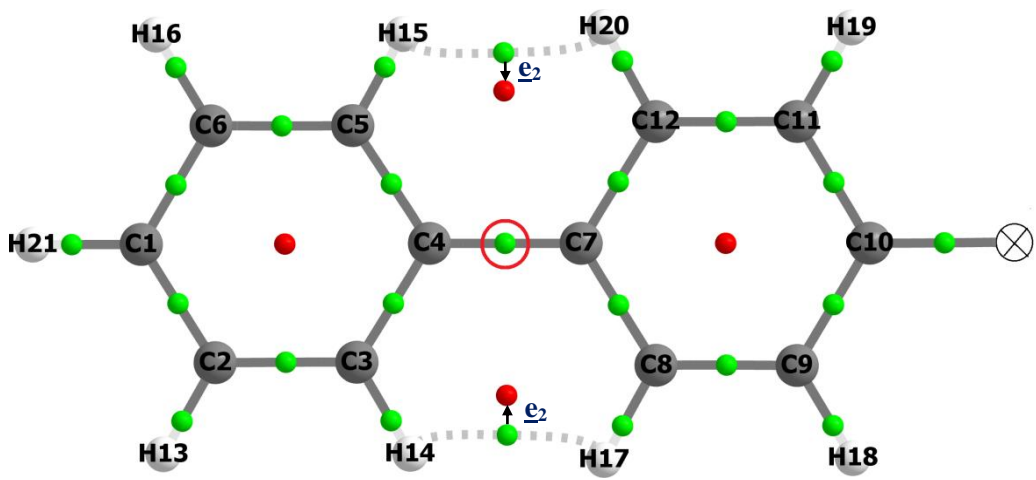

**Scheme 2**. The representation of the molecular graphs of the *para*-substituted biphenyl that undergo the torsion θ, $C_{12}H_9$-*x* where *x* = $N(CH_3)_2$, $NH_2$, $CH_3$, CHO, CN, $NO_2$. The C1-C2-C3-C4-C5-C6 ring is the fixed reference ring. The C7-C8-C9-C10-C11-C12 ring is rotated so that the nuclei comprising the substituent groups move out of the plane in an anti-clockwise direction and the H20 nuclei moves into the plane of the C1-C2-C3-C4-C5-C6 ring. The atom labels correspond to the molecular graph atom numbering scheme and the torsion C4-C7 *BCP* is indicated by a red circle. The directions of the **e₂** eigenvectors for the H---H *BCP*s are indicated by the blue arrows and lie in the plane of the biphenyl molecular graph for torsion θ = 0.0°.

The first step of the computational protocol is to perform a relaxed scan of the potential energy surface (PES) of the torsion C4-C7 *BCP*. The PES was performed at 1.0° intervals over 0.0° ≤ θ ≤ 180.0° except for 0.0° ≤ θ ≤ 10.0° where it was performed at 0.2° intervals with a geometry optimization performed at all steps, spanning the reaction coordinate θ; 0.0° ≤ θ < 25.0°, with tight convergence criteria at B3LYP/6-311G(2d,3p) with Gaussian 09B01[17]. Subsequent single point energies for each step in the PES were evaluated at the same theory level. Then the QTAIM analysis was performed with the AIMAll[18] suite on each wave function obtained in the previous step. The procedure to generate the eigenvector-following path lengths $\mathbb{H}$ and $\mathbb{H}^*$ is provided in **Supplementary Materials S1**.

For the purposes of comparison with the relative energy ΔE results as well as the eigenvector following path lengths $\mathbb{H}$ and $\mathbb{H}^*$ we include the variation of the bond-path lengths (BPLs) with torsion θ, of the torsion C4-C7 *BCP* and the H15---H20 *BCP* for the *para*-substituted biphenyl $C_{12}H_9$-x where x = $N(CH_3)_2$, $NH_2$, $CH_3$, CHO, CN, $NO_2$, see the black plots in each of the sub-figures (**a-d**) of **Figure 1**, also see **Scheme 2**. The black plots; the relative energy ΔE for the undecorated biphenyl[19] and the BPLs were previously published by some of the current authors[4] but we include here as the BPL is the length of the *r* path that comprises the bond-path framework set $\mathbb{B} = \{p, q, r\}$. It can be seen that the form of the variation of the relative energy ΔE with the torsion θ is closer to that of the eigenvector-following path lengths $\mathbb{H}$ than that of the corresponding variation of the BPL, compare **Figure 1(a)** with **Figure 1(b)**, where the black plots in **Figure 1(b)** correspond to the variation of the BPL. In particular, we see a greater spread of values of the relative energy ΔE for the *para*-substituted groups for values of the torsion θ = 90.0° than in the vicinity of the biphenyl energy minimum θ ≈ 40.0°. In addition, the position of the minimum values of $\mathbb{H}$ in the vicinity of θ ≈ 40.0° more closely resemble the corresponding distributions of the relative energy ΔE than do the BPL that possess a wider spread of values.

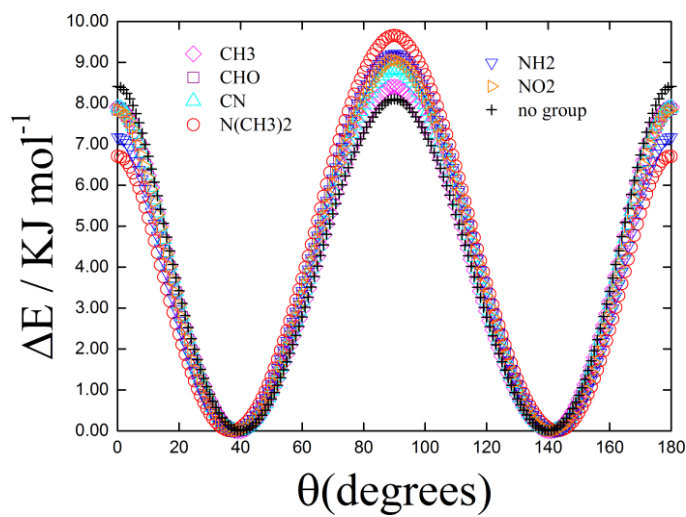

(a)

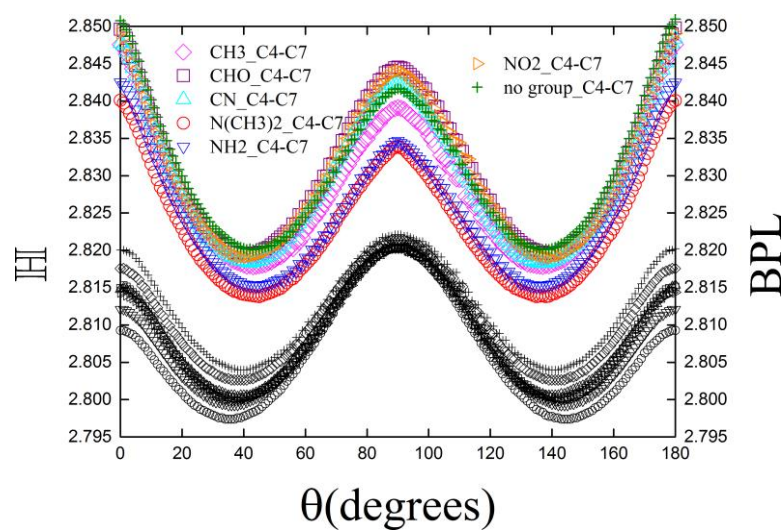

(b)

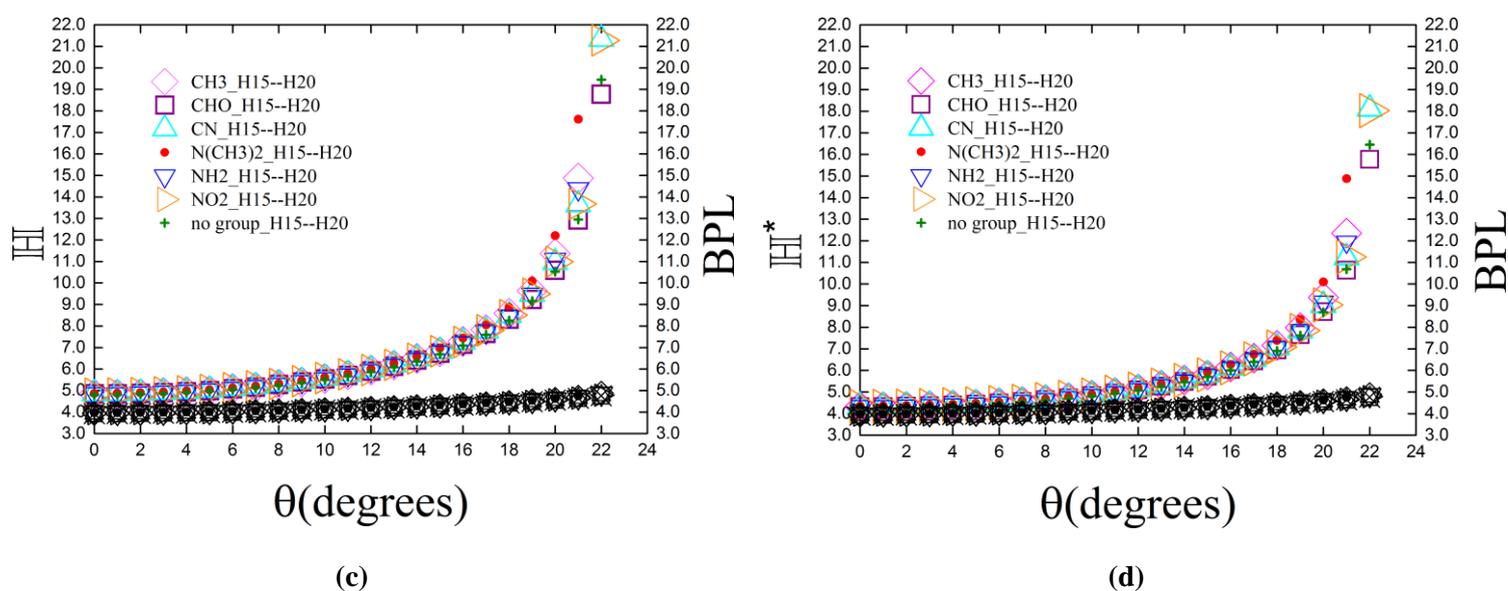

**(c)** **(d)**

**Figure 1**. The relative energy ΔE of the torsion θ for the *para*-substituted biphenyl is shown in sub-figures **(a)**, the variations of the eigenvector-following path length $\mathbb{H}$ and BPL of the C4-C7 *BCP* with torsion θ are denoted by the multi-colored and black plots respectively in sub-figure **(b)**, the corresponding values for $\mathbb{H}^*$ are provided in **Figure S2(a)** of the **Supplementary Materials S2**. The values of $\mathbb{H}$ and $\mathbb{H}^*$ of the H15---H20 *BCP* are shown in sub-figures **(c)-(d)** respectively, the corresponding values for $\mathbb{H}$, $\mathbb{H}^*$ of the H14---H17 *BCP* are provided in **Figure S2(b-c)** of the **Supplementary Materials S2**.

For the torsion C4-C7 *BCP* the values of lengths $\mathbb{H}$ and $\mathbb{H}^*$ are indistinguishable due to the lack of significant bond-path curvature, compare **Figure 2(a)** with **Figure S2(a)** of the **Supplementary Materials S2**. In contrast to the torsion C4-C7 *BCP*, the H15---H20 *BCP* exhibits significant non-zero bond-path curvature, compare **Figure 2(a)** and **Figure 2(b-c)** respectively. The values of the lengths $\mathbb{H}$ for a given value of the torsion θ always exceed those of the BPL, compare the colored plots with the black plots in **Figure 1(b)**. The length $\mathbb{H}$ exceed that of the BPL (path *r*) and can be seen from of the form of the path $q_i = r_i + \varepsilon_i \underline{e}_{2,i}$, where non-zero ellipticity ε values of the C4-C7 *BCP* result in values of the length $\mathbb{H}$ > BPL. For a particular *para*-substituted group we see that variation of the $\mathbb{H}$ lengths with torsion θ resembles the corresponding variation in the relative energy ΔE. This is due to shorter $\mathbb{H}$ lengths corresponding to lower ellipticity ε values from the form of $q_i = r_i + \varepsilon_i \underline{e}_{2,i}$, and lower ellipticity ε values offer less resistance to an applied torsion θ which results in a lower relative energy ΔE.

Examination of the H15---H20 *BCP* that connects the two phenyl rings, see **Scheme 2**, ruptures for at an applied torsion θ ≈ 22.0°, see **Figure 1(c-d).** The variations of $\mathbb{H}$ and $\mathbb{H}^*$ with the applied torsion θ are indicated with colored plots and the corresponding variation of the BPL is indicated by black plots. The changes to the lengths $\mathbb{H}$ and $\mathbb{H}^*$ shown by the colored plots, are much more apparent in comparison with BPL, both in terms of greater separation of the values of the *para*-substituted groups and increase in the overall $\mathbb{H}$ and $\mathbb{H}^*$ lengths compared with the BPLs. At values of the torsion θ ≥ 17.0° the values for $\mathbb{H}$ and also for $\mathbb{H}^*$ the *para*-substituted groups separate, this contrasts with the behavior the BPL for the *para*-substituted groups that do not separate for any value of torsion θ. As a consequence the $\mathbb{H}$ and $\mathbb{H}^*$ lengths will distinguishable as can be seen in **Figure 1(b-c)**, where the values of $\mathbb{H} > \mathbb{H}^*$ for all values of the applied torsion θ before the rupture of the H---H *BCP*. This is due to the $\mathbb{H}$ and $\mathbb{H}^*$ being constructed from the most ($\underline{\mathbf{e}}_2$) and least ($\underline{\mathbf{e}}_1$) preferred directions of the charge density ρ(**r**) respectively therefore $\mathbb{H}$ will sweep out a longer path than $\mathbb{H}^*$. This is explained in **Scheme 2**, where we see that bonding bends in direction of the $\underline{\mathbf{e}}_2$ eigenvector of the H---H *BCP*s. For values of θ > 0.0° the H---H *BCP*s begin to slide in the most preferred $\underline{\mathbf{e}}_2$ direction towards with the closest *RCP*, shown as the undecorated red sphere, before coalescence and annihilation at θ ≈ 22.0°.

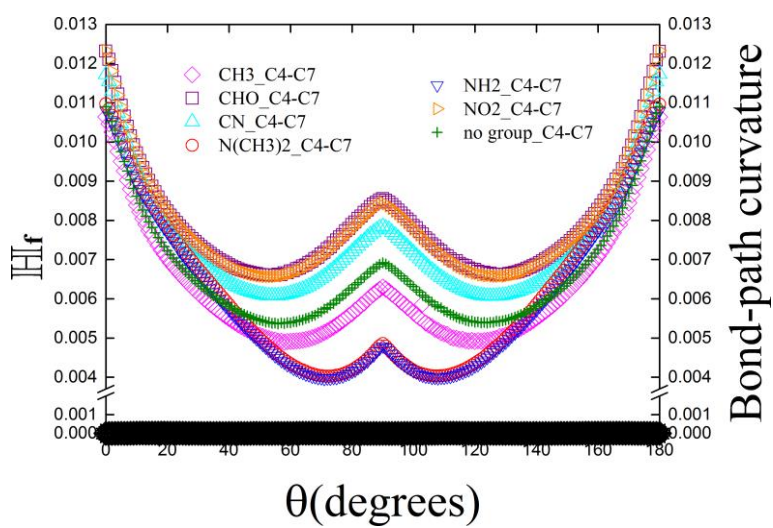

(a)

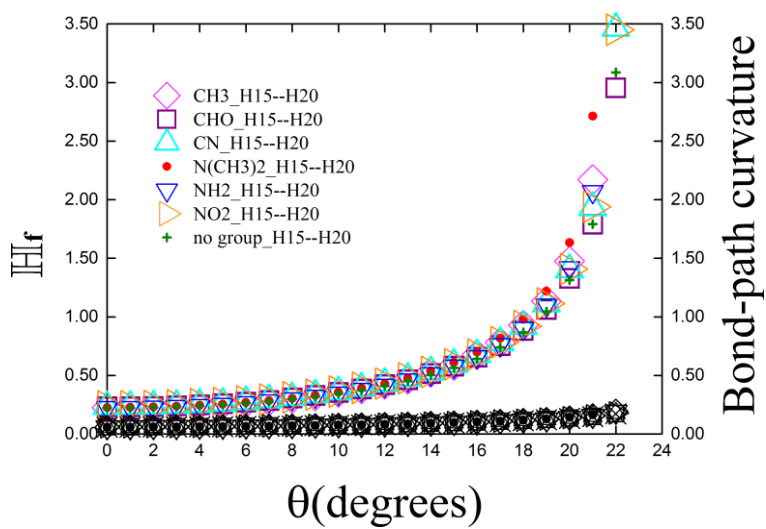

(b)

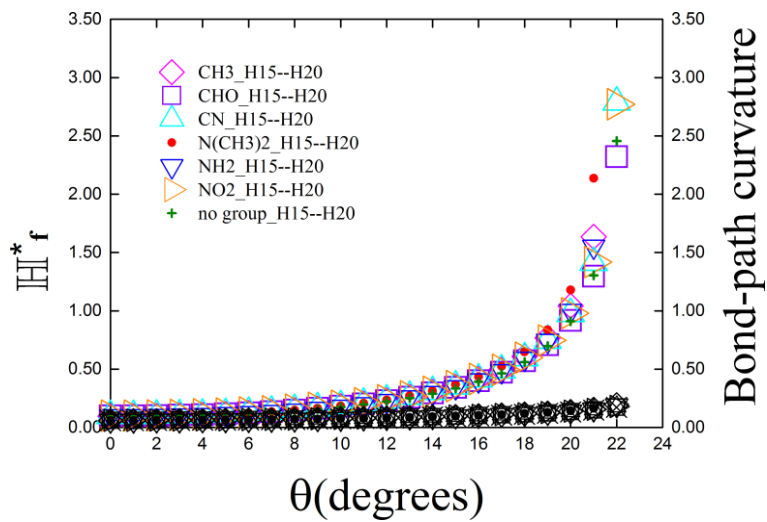

(c)

**Figure 2**. The variation of the fractional eigenvector-following path length $\mathbb{H}_f$ and the bond-path curvature (BPL - GBL)/GBL, see equation **(1)**, of the C4-C7 *BCP* with torsion θ are denoted by the multi-colored and black plots respectively in sub-figure **(a)**, the corresponding values for $\mathbb{H}^*_f$ are provided in **Figure S3(a)** of the **Supplementary**

**Materials S3.** The values of $\mathbb{H}_f$ and $\mathbb{H}^*_f$ of the H15---H20 *BCP* are shown in sub-figures **(b)-(c)** respectively, the corresponding values for $\mathbb{H}_f$, $\mathbb{H}^*_f$ of the H14---H17 *BCP* are provided in **Figure S3(b-c)** of the **Supplementary Materials S3**.

We examine the first fractional eigenvector following path lengths $\mathbb{H}_f$ for the *para*-substituted biphenyl and find them to be significant for the both the torsion C4-C7 *BCP* and the H15---H20 *BCP* and orders of magnitude larger than the equivalent fractional measure for the BPL, the bond-path curvature, see **Figure 2**. This is due to the torsional twisting distortion disrupting the relative orientation of the $\underline{e_1}$ and $\underline{e_2}$ eigenvectors relative to the $\underline{e_3}$ eigenvector orders of magnitude more than the bending distortion of the bond-path of C4-C7 *BCP* that effects the bond-path curvature. Similar trends are seen for the other fractional measures $\mathbb{H}_{f\theta min}$ and $\mathbb{H}^*_{f\theta min}$, see **Figure 3**. We notice that the values of $\mathbb{H}_f$ and $\mathbb{H}^*_f$, although distinguishable from one another, of the H15---H20 *BCP* are indistinguishable from the corresponding $\mathbb{H}_{f\theta min}$ and $\mathbb{H}^*_{f\theta min}$ plots, compare **Figure 2(b-c)** with **Figure 3(b-c)**. Conversely, for the torsion C4-C7 *BCP* the two fractional measures $\mathbb{H}_f$ and $\mathbb{H}_{f\theta min}$ are very different, as seen from the comparison of **Figure 2(a)** with **Figure 3(a)**.

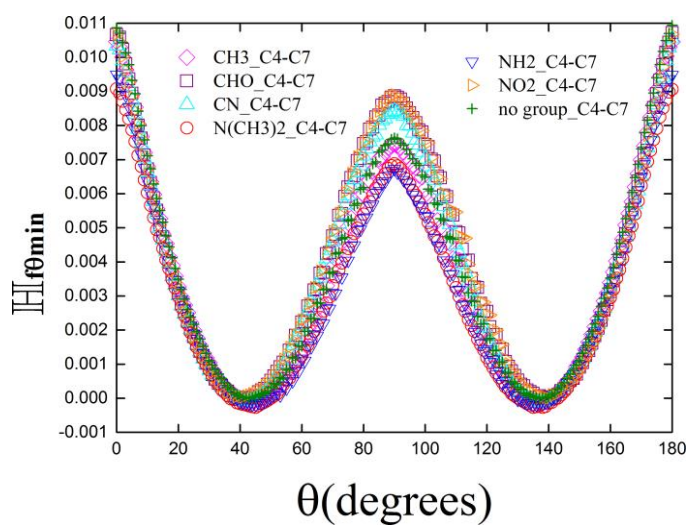

(a)

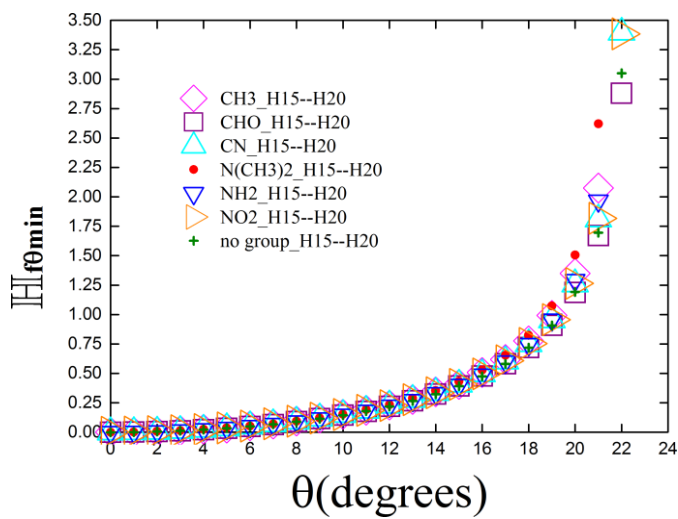

(b)

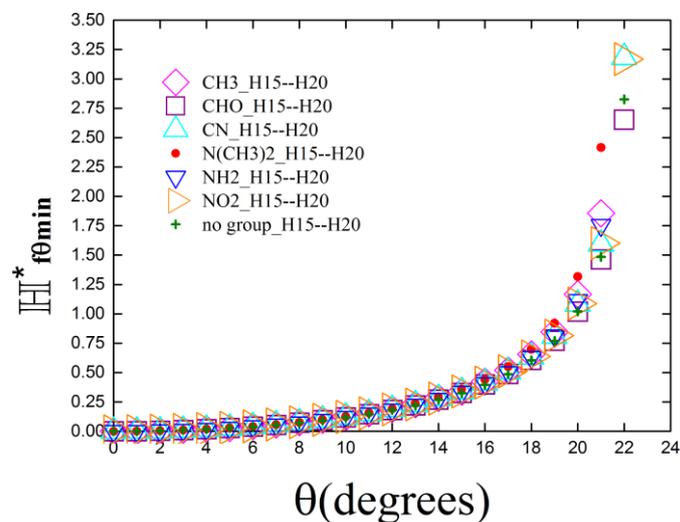

(c)

**Figure 3.** The variation of the fractional eigenvector-following path length $\mathbb{H}_{f\theta min} = (\mathbb{H} - \mathbb{H}_{\theta min})/\mathbb{H}_{\theta min}$ of the C4-C7

*BCP* with torsion θ is shown in sub-figure **(a)**, the corresponding values for $\mathbb{H}^*_{f\theta min}$ are provided in **Figure S4(a)** of the **Supplementary Materials S4**. The values of $\mathbb{H}_{f\theta min}$ and $\mathbb{H}^*_{f\theta min}$ of the H15---H20 *BCP* are shown in sub-figures **(b)-(c)** respectively, the corresponding values for $\mathbb{H}_{f\theta min}$, $\mathbb{H}^*_{f\theta min}$ of the H14---H17 *BCP* are provided in **Figure S4(b-c)** of the **Supplementary Materials S4**.

To summarize, we have applied the new vector-based interpretation of the chemical bond, the bond-path framework set $\mathbb{B} = \{p, q, r\}$ to the molecular graph of a *para*-substituted biphenyl molecule $C_{12}H_9$- *x* where *x* = $N(CH_3)_2$, $NH_2$, $CH_3$, CHO, CN, $NO_2$ subjected to a torsion θ over the range $0.0° \leq \theta \leq 180.0°$. The paths *p*, *q*, *r* have lengths $\mathbb{H}^*$, $\mathbb{H}$ and the bond-path length (BPL) respectively that are defined in terms of the eigenvectors $\underline{e}_1$, $\underline{e}_2$ and $\underline{e}_3$ of the Hessian matrix respectively. The lengths of the *p* and *q* paths given by $\mathbb{H}^*$, $\mathbb{H}$ increase significantly more than occurs for the variation of the BPL with applied torsion θ for the torsion C4-C7 *BCP* and the H---H *BCP*s. This difference occurs as a consequence of insensitivity of the BPL in measuring the effect on a bond-path of an applied torsion θ for all choices of *x* for the *para*-substituent group. As expected, the applied torsion θ does not significantly increase the length of the bond-path but instead affects the relative orientation of the *p* and *q* paths relative to the *r* path, in other words the bond-path twists but does not stretch significantly. This clearly demonstrates the importance of the inclusion of the $\underline{e}_1$ and $\underline{e}_2$ eigenvectors into any definition of the chemical bond in addition to the $\underline{e}_3$ eigenvector particularly when a torsion distortion is present. Realistic distortions of a chemical bond during in a chemical reaction will involve a degree of torsion in addition to stretching/compression distortions where the later only affects the BPL i.e. the $\underline{e}_3$ eigenvector.

We also saw that for the H---H *BCP* that the lengths of $\mathbb{H} > \mathbb{H}^*$ was due to the lengths $\mathbb{H}$ and $\mathbb{H}^*$ being constructed from the most ($\underline{e}_2$) and least ($\underline{e}_1$) preferred directions of the charge density $\rho(\mathbf{r})$ respectively. Additionally, we demonstrated the uniqueness of both of the new fractional measures $\mathbb{H}_{f\theta min}$ and $\mathbb{H}^*_{f\theta min}$ for the consideration of the strong torsion C4-C7 *BCP*.

Currently the explicit plotting the *p* and *q* paths for both ground state and excited state calculations is being implemented.

**Acknowledgements**

The National Natural Science Foundation of China is gratefully acknowledged, project approval number: 21673071. The One Hundred Talents Foundation of Hunan Province and the aid program for the Science and Technology Innovative Research Team in Higher Educational Institutions of Hunan Province are also gratefully acknowledged for the support of S.J. and S.R.K.

# SUPPLEMENTARY MATERIALS

# A Vector-Based Representation of the Chemical Bond for the Substituted Torsion of Biphenyl

**Jiahui Li, Tianlv Xu, Steven R. Kirk*  and Samantha Jenkins***

[1]*Key Laboratory of Chemical Biology and Traditional Chinese Medicine Research and Key Laboratory of Resource Fine-Processing and Advanced Materials of Hunan Province of MOE, College of Chemistry and Chemical*
[2]*Engineering, Hunan Normal University, Changsha, Hunan 410081, China*
[2]*EaStCHEM School of Chemistry, University of St Andrews, North Haugh, St Andrews, Fife KY16 9ST, Scotland, United Kingdom.*

*email: samanthajsuman@gmail.com
*email: steven.kirk@cantab.net

**1. Supplementary Materials S1.** Implementation details of the calculation of the eigenvector-following path lengths $\mathbb{H}$ and $\mathbb{H}^*$.

**2. Supplementary Materials S2.** The variation of the eigenvector-following path length $\mathbb{H}$ with the torsion θ of the H14--H17 *BCP*, the values for $\mathbb{H}^*$ of the C4-C7 *BCP* and H14--H17 *BCP* for the substituted biphenyl.
.

**3. Supplementary Materials S3.** The variation of the eigenvector-following path length $\mathbb{H}_f$ with the torsion θ of the H14--H17 *BCP*, the values for $\mathbb{H}^*_f$ of the C4-C7 *BCP* and H14--H17 *BCP* for the substituted biphenyl.
.

**4. Supplementary Materials S4.** The variation of the $\mathbb{H}_{fθmin}$ with the torsion θ of the H14--H17*BCP*, the corresponding values for $\mathbb{H}^*_{fθmin}$ of the C4-C7 *BCP* and H14--H17 *BCP* for the substituted biphenyl.

**1. Supplementary Materials S1.** Implementation details of the calculation of the eigenvector-following path lengths $\mathbb{H}$ and $\mathbb{H}^*$.

When the QTAIM eigenvectors of the Hessian of the charge density $\rho(\mathbf{r})$ are evaluated at points along the bond-path, this is done by requesting them via a spawned process which runs the selected underlying QTAIM code, which then passes the results back to the analysis code. For some datasets, it occurs that, as this evaluation considers one point after another in sequence along the bond-path, the returned calculated $\underline{\mathbf{e}}_2$ (correspondingly $\underline{\mathbf{e}}_1$ is used to obtain $\mathbb{H}^*$) eigenvectors can experience a 180-degree 'flip' at the 'current' bond-path point compared with those evaluated at both the 'previous' and 'next' bond-path points in the sequence. These 'flipped' $\underline{\mathbf{e}}_2$ (or $\underline{\mathbf{e}}_1$) eigenvectors, caused by the underlying details of the numerical implementation in the code that computed them, are perfectly valid, as these are defined to within a scale factor of -1 (i.e. inversion). The analysis code used in this work detects and re-inverts such temporary 'flips' in the $\underline{\mathbf{e}}_2$ (or $\underline{\mathbf{e}}_1$) eigenvectors to maintain consistency with the calculated $\underline{\mathbf{e}}_2$ (or $\underline{\mathbf{e}}_1$) eigenvectors at neighboring bond-path points, in the evaluation of eigenvector-following path lengths $\mathbb{H}$ and $\mathbb{H}^*$.

## 2. Supplementary Materials S2.

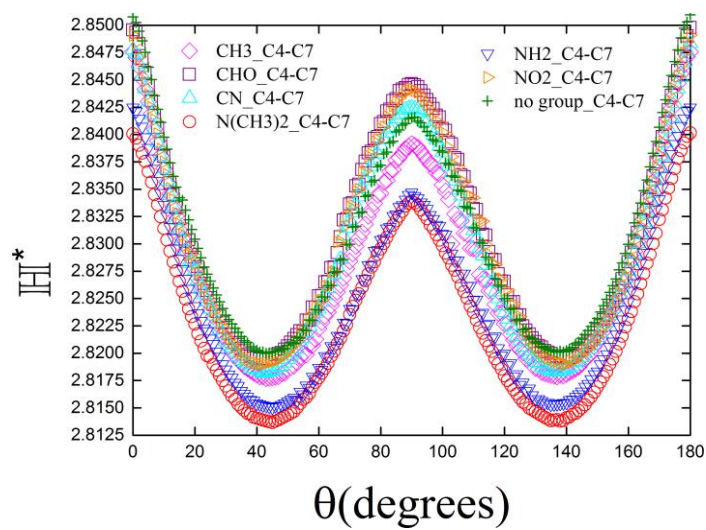

(a)

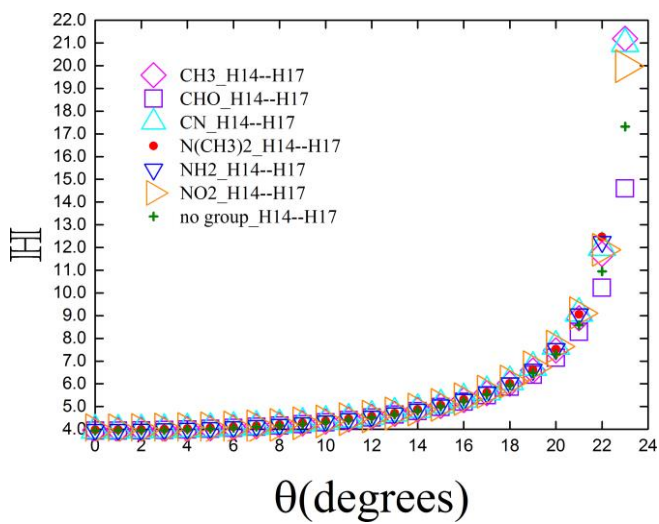

(b)

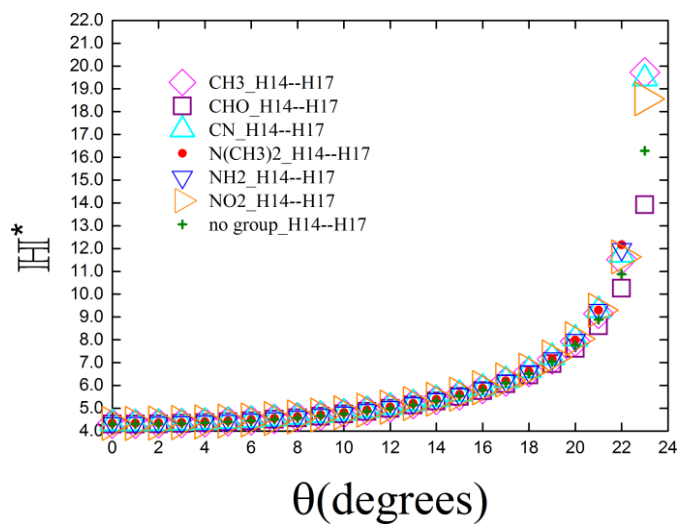

(c)

**Figure S2.** The variation of the eigenvector-following path length $\mathbb{H}^*$ of the C4-C7 *BCP* with torsion θ is shown in sub-figure **(a)**, the values of $\mathbb{H}$ and $\mathbb{H}^*$ of the H14--H17 *BCP* are shown in sub-figures **(b)-(c)** respectively.

## 3. Supplementary Materials S3.

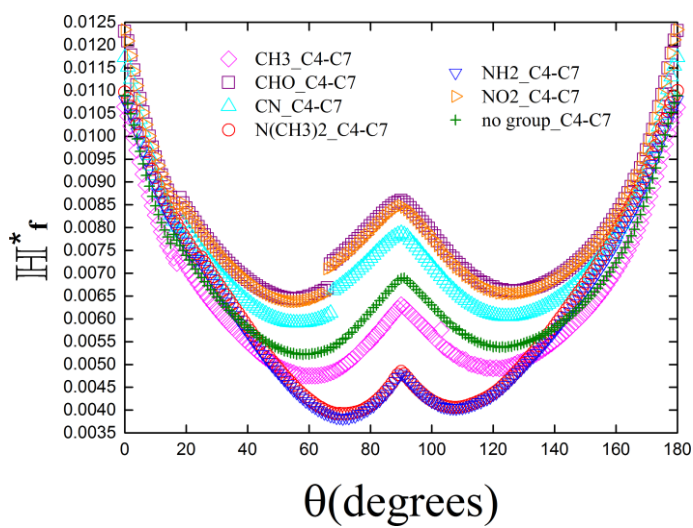

(a)

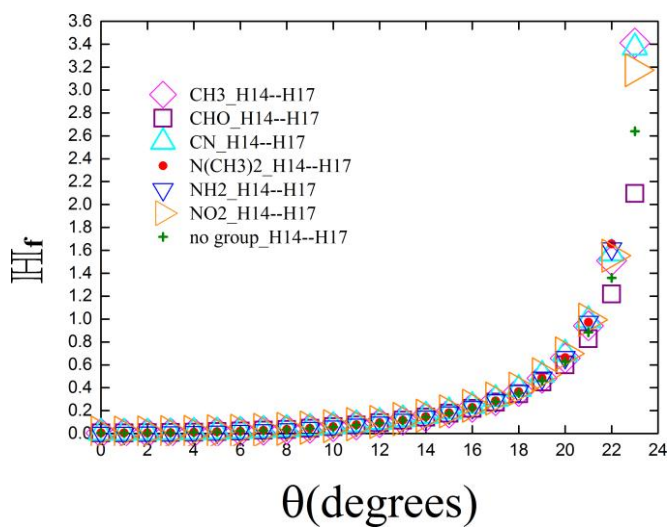

(b)

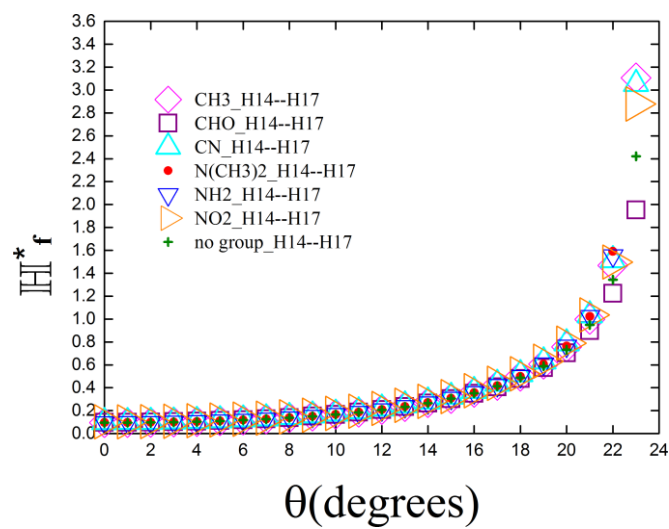

(c)

**Figure S3.** The variation of the fractional eigenvector-following path length $\mathbb{H}^*_f$ of the C4-C7 *BCP* with torsion $\theta$ are presented in sub-figure **(a)**, the corresponding values for $\mathbb{H}_f$ and $\mathbb{H}^*_f$ of the H14--H17 *BCP* are presented in sub-figures **(b)-(c)** respectively.

## 4. Supplementary Materials S4.

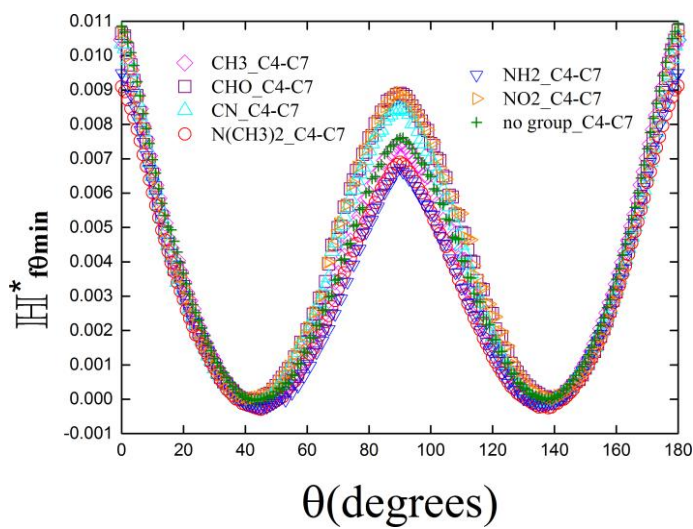

(a)

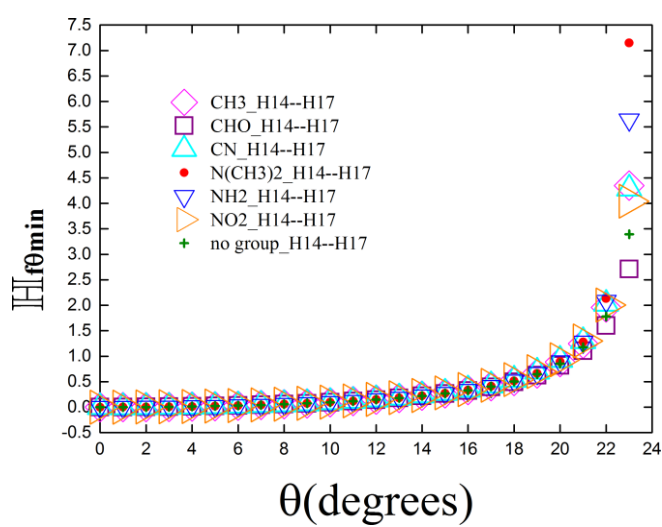

(b)

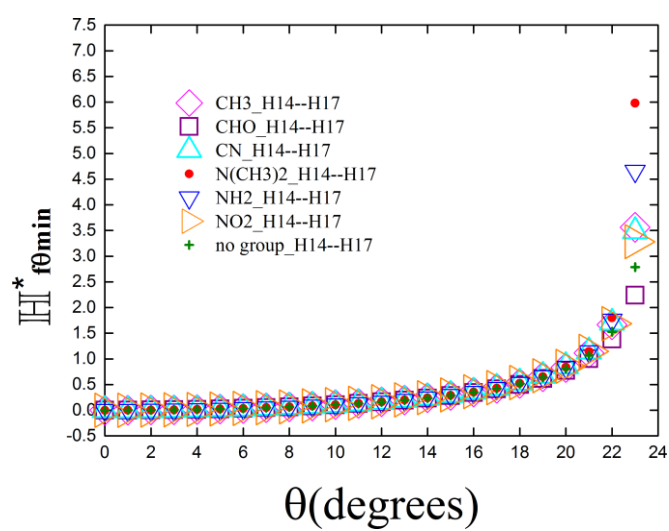

(c)

**Figure S4.** The variation of the fractional eigenvector-following path length $\mathbb{H}^*_{f\theta min} = (\mathbb{H}^* - \mathbb{H}^*_{\theta min})/\mathbb{H}^*_{\theta min}$ of the C4-C7 *BCP* with torsion θ is shown in sub-figure **(a)**, the values of $\mathbb{H}_{f\theta min}$ and $\mathbb{H}^*_{f\theta min}$ of the H14--H17 *BCP* are shown in sub-figures **(b)-(c)** respectively.